\documentclass[twocolumn,showpacs,preprintnumbers,amsmath,amssymb]{revtex4}
\usepackage{graphicx}

\begin{document}

\thispagestyle{empty}
\noindent
{\bf
Comment on ``Casimir Force and {\it In Situ} Surface Potential
Measurements on Nanomembranes''
}

In Ref.~\cite{1} the frequency shift of an oscillator
$\Delta f$ under the
influence of the residual electric force $F_{\rm res}^{\rm el}(z)$
and the Casimir force $F_C(z)$ was measured between Au-coated
surfaces of a sphere and a membrane. Using the model for
$F_{\rm res}^{\rm el}(z)$ with two fitting parameters it was
claimed that the data for frequency shift are best described by
the Drude model approach to the Casimir
force ($\chi^2=35.3$ and $\chi^2$-probability to
exceed 35\%) and excludes the plasma model approach
($\chi^2=56.1$ and the probability of about 1\%).
We demonstrate that these results are incorrect, and that the
data of Ref.~\cite{1} are
 inconsistent with both theoretical approaches.

First, we note that although a mistake in
Eq.~(2) in Ref.~\cite{1}, indicated by us earlier \cite{com},
was corrected in an Erratum \cite{er},
the factor $\eta=\sqrt{1+A_{\rm rms}^2/z^2}$, where $A_{\rm rms}$
is the r.m.s. amplitude of membrane vibrations
and $z$ is separation, remained incorrect.
In dynamic experiments the corrections due to surface roughness
must be included in an expression for the external force,
whereas vibrations determine the frequency shift according to
corrected Eq.~(2). Because of this, the factor $\eta$ used on
p.4 to correct separations must be replaced with
$\eta_{\rm corr}=\sqrt{1+3A_{\rm rms}^2/(2z^2)}$.

Next, we recalculated $\Delta f$ using
the tabulated optical data of Au \cite{2} extrapolated to zero
frequency using either the Drude \cite{BS}
or the plasma \cite{4} model with
parameters $\omega_p=7.54\,$eV and $\gamma=0.051\,$eV which,
according to Refs.~\cite{1,3}, best fit the optical data.
In Fig.~1, our computational results for
$z\Delta f$ using corrected Eq.~(2) and the Drude- and
plasma-model extrapolations are shown by the upper and lower solid
lines, respectively, for $z$ from 0.118
to $0.230\,\mu$m where, as stated in Ref.~\cite{1}, the
electrostatic force is negligible. The experimental data
taken from
Fig.~4(c) of Ref.~\cite{1} are indicated as crosses whose arms
show the experimental errors (the values of $z$ were extracted
from the electrostatic force and corrected by a factor $\eta$
in the Letter \cite{1}).

As is seen in Fig.~1, all 15
data points of the total 32 exclude both the Drude- and
plasma-model approaches to the Casimir force.
For any fit, the contribution of these data to $\chi^2$ will
exceed 300 and 419, respectively. The 17 data points,
measured in Ref.~\cite{1} at larger separations, where
$F_{\rm res}^{\rm el}(z)$ may be not negligible, can only increase
these values. As a result, for both approaches the
$\chi^2$-probability is much less than $10^{-6}\,$\%.
Our computational results do not coincide with the theoretical
results in Fig.~4(c) of Ref.~\cite{1}. The latter are reproduced
when one disregards the optical data and uses
instead simple
Drude and plasma models over the entire frequency region and
Eq.~(2) with only the first term
(see the upper and lower dashed lines on an
inset to Fig.~1, respectively).
It is known, however, that at $z\lesssim 0.3\,\mu$m
core electrons contribute significantly to the dielectric
permittivity, so that simple Drude and plasma models cannot
be used to calculate the Casimir force \cite{4}.
Note that the effect
of surface roughness only increases the
magnitude of $\Delta f$, thus, increasing a disagreement
between experiment and theory.

To summarize, the calculations presented here demonstrate
strong disagreement between
the experimental data of Ref.~\cite{1} and both theoretical approaches to the Casimir
force, thus, suggesting the presence of
an unaccounted systematic error
in the data.

This work was supported by the DFG grant BO\ 1112/21--1.

\noindent
M.~Bordag,${}^1$ G.~L.~Klimchitskaya,${}^{1,2}$
and V.\ M.\ Mos\-te\-pa\-nen\-ko${}^{1,2}$ \hfill \\
${}^1$Institute for Theoretical
Physics, Leipzig University,
D-04009, Leipzig, Germany \hfill \\
${}^2$Central Astronomical Observatory
at Pulkovo of the Russian Academy of Sciences,
St.Petersburg, 196140, Russia
\hfill \\[2mm]

\begin{figure}[h]
\vspace*{-13.1cm}
\centerline{
\hspace*{-0.5cm}
\includegraphics{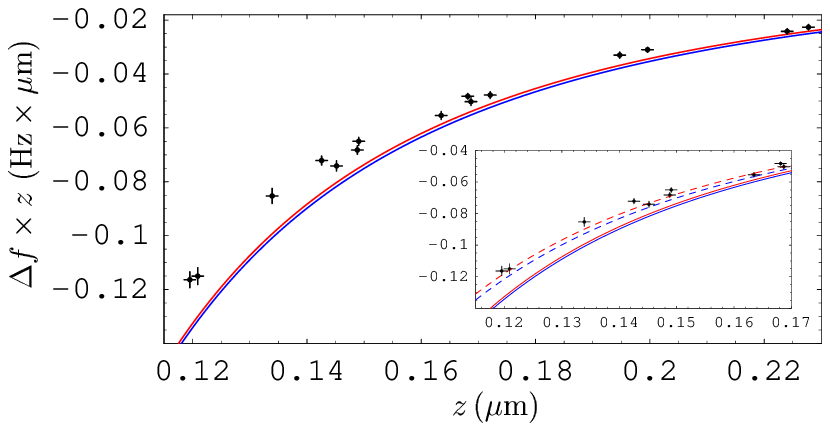}
}
\vspace*{-12.8cm}
\caption{(Color online)
The experimental data with respective errors are indicated as
crosses. The upper and lower solid lines (dashed lines) show
the theoretical results calculated here using the optical
data extrapolated by means of the Drude and plasma models,
respectively (calculated in Ref.~\cite{1} using the Drude and
plasma models).}
\end{figure}


\end{document}